\documentclass{llncs}
\usepackage{amsmath}
\usepackage{amssymb}
\usepackage{enumerate}

\newcounter{instr}
\newcommand{\ninstr}{\refstepcounter{instr}\theinstr.}
\newcommand{\SP}{((p_1, q_1), (p_2, q_2))}
\newcommand{\klcs}[1]{$#1$-LCF}
\newcommand{\pos}{p}

\title{Longest common substrings with $k$ mismatches}

\author{Tomas Flouri\inst{1} \and Emanuele Giaquinta\inst{2} \and Kassian Kobert\inst{1} \and Esko Ukkonen\inst{3}}

\institute{Heidelberg Institute for Theoretical Studies, Germany\\
          \email{\{Tomas.Flouri $\mid$ Kassian.Kobert\}@h-its.org} \and
          Department of Computer Science, Aalto University, Finland
          \email{emanuele.giaquinta@aalto.fi} \and
          Department of Computer Science, University of Helsinki, Finland
          \email{ukkonen@cs.helsinki.fi}}

\begin{document}

\maketitle

\begin{abstract}
The longest common substring with $k$-mismatches problem is to find,
given two strings $S_1$ and $S_2$, a longest substring $A_1$ of $S_1$
and $A_2$ of $S_2$ such that the Hamming distance between $A_1$ and
$A_2$ is $\le k$. We introduce a practical $O(nm)$ time and $O(1)$
space solution for this problem, where $n$ and $m$ are the lengths of
$S_1$ and $S_2$, respectively. This algorithm can also be used to
compute the matching statistics with $k$-mismatches of $S_1$ and $S_2$
in $O(nm)$ time and $O(m)$ space. Moreover, we also present a
theoretical solution for the $k = 1$ case which runs in $O(n \log m)$
time, assuming $m\le n$, and uses $O(m)$ space, improving over the
existing $O(nm)$ time and $O(m)$ space bound of Babenko and
Starikovskaya~\cite{Babenko11}.
\end{abstract}

\section{Introduction}
In this paper we study the longest common substring (or \emph{factor}) with
$k$-mismatches problem (\klcs{k} for short\footnote{We use the \klcs{k}
abbreviation as LCS usually refers to the \emph{Longest Common Subsequence}
problem}) which consists in finding the longest common substring of two strings
$S_1$ and $S_2$, while allowing for at most $k$ mismatches, i.e., the Hamming
distance between the two substrings is $\le k$.
This problem is a generalization of the Longest Common Substring problem~\cite{Gusfield1997,StarikovskayaV13,KociumakaSV14}
and is similar to the \emph{threshold all-against-all} problem defined by
Gusfield~\cite{Gusfield1997} and to the \emph{local alignment} problem of biological sequence analysis. In the
threshold all-against-all problem the goal is to find all the pairs of
substrings of $S_1$ and $S_2$ such that the corresponding edit distance is less
than a given number $d$. The difference in the \klcs{k} problem is that the
distance used is the Hamming distance rather than the edit distance, and that
we are interested in the pairs of substrings of maximal length only. In the
local alignment problem, which can be solved in $O(|S_1|\cdot |S_2|)$ time using the
Smith-Waterman algorithm~\cite{SmithW81}, the goal is to compute a pair of
substrings of $S_1$ and $S_2$ such that the corresponding similarity, according
to a suitable scoring function, is maximum over all the pairs of substrings. In
particular, if the scoring function is such that the score of a match is $1$,
the score of a mismatch is $0$ and gaps are not allowed, a solution of the
local alignment problem is comparable to one of the \klcs{k} problem, with the
difference that there is no bound on the number of mismatches.

Babenko and Starikovskaya~\cite{Babenko11} studied the case of $1$ mismatch only and
presented an algorithm for the \klcs{1} problem which runs in $O(|S_1|\cdot
|S_2|)$ time.  A closely related problem is the one of computing the matching
statistics with $k$ mismatches. The matching statistics, introduced by Chang
and Lawler~\cite{ChangL94} for the approximate string matching problem, is an
array $ms$ of $|S_2|$ integers such that $ms[i]$ is the length of the longest
substring of $S_2$ that starts at position $i$ and matches exactly
some substring of $S_1$, for $i = 0, \ldots, |S_2|-1$. A natural generalization
is obtained by allowing the matching to be approximate, with
respect to the Hamming distance.  Recently, Leimeister and
Morgenstern~\cite{LeimeisterM14} presented a greedy heuristic for the
computation of the matching statistics with $k$ mismatches, which runs in
$O(|S_1|\cdot k\cdot z)$ time, where $z$ is the maximum number of occurrences
in $S_2$ of a string of maximal length which occurs in both $S_1$ and $S_2$.

In this paper we present two novel contributions. Our first result is
an efficient algorithm for the \klcs{k} problem which runs in time
$O(|S_1|\cdot |S_2|)$ and only requires a constant amount of space.
This algorithm can also be used to compute the matching statistics
with $k$ mismatches with no overhead in the time complexity, i.e., in
$O(|S_1|\cdot |S_2|)$ time, and using $O(|S_2|)$ space. Our second
result is an algorithm for the \klcs{1} problem, i.e., for the $k=1$
case. We show how to solve this instance in a more time efficient
manner by using results from Crochemore et al.~\cite{CrochemoreIMS06}
for finding the longest generalized repeat(s) with one block of $k$
adjacent don't care symbols. Assuming $|S_2| \le |S_1|$, our algorithm
takes time $O(|S_1|\log |S_2|)$, improving over the previous bound of
$O(|S_1| \cdot |S_2|)$.

\section{Basic definitions}
Let $\Sigma$ be a finite alphabet of symbols and let $\Sigma^*$ be the
set of strings over $\Sigma$. Given a string $S\in\Sigma^*$, we denote by $|S|$
the length of $S$ and by $S[i]$ the $i$-th symbol of $S$, for $0\le
i < |S|$. Given two strings $S$ and $S'$, $S'$ is a substring of $S$
if there are indices $0\le i \leq j < |S|$ such that $S' = S[i] ... S[j]$.
If $i = 0$ ($j = |S| - 1$) then $S'$ is a prefix (suffix) of $S$. We
denote by $S[i .. j]$ the substring of $S$ starting at position $i$
and ending at position $j$. For $i > j$ we obtain the empty string
$\varepsilon$. Finally, we denote by $S^r = S[|S| - 1] S[|S|-2] \ldots S[0]$ the reverse of the string $S$.

The suffix tree $\mathcal{T}(S)$ of a string $S$ is a rooted directed
tree with $|S'|$ leaves and edge labels over
$(\Sigma\cup\{\$\})^*\setminus\{\varepsilon\}$, where
$\$\notin\Sigma$ and $S' = S\$$. Each internal node has at least two children and is
such that the edge labels of the children have different first
symbols. For each leaf $i$, the concatenation of the edge labels on
the path from the root to leaf $i$ is equal to $S'[i .. |S'| - 1]$.
Assuming a constant size alphabet, the suffix tree can be built in
$O(|S|)$ time~\cite{Gusfield1997}. For any node $u$ in
$\mathcal{T}(S)$, $depth(u)$ denotes the length of the string labeling
the path from the root to $u$. For any pair of nodes $u,v$ in
$\mathcal{T}(S)$, $LCA(u, v)$ denotes the lowest common ancestor of
$u$ and $v$, i.e., the deepest node in $\mathcal{T}(S)$ that is
ancestor of both $u$ and $v$. The suffix tree can be preprocessed in
$O(|S|)$ time so as to answer LCA queries in constant
time~\cite{BenderF00}. We denote by $\mathcal{B}(S)$ the binary
suffix tree obtained by replacing each node $u$ in $\mathcal{T}(S)$
with out-degree at least $2$ with a binary tree with $d-1$
internal nodes (whose $depth$ values are equal to $depth(u)$) and $d-2$ internal edges, where the $d$ leaves are the
$d$ children of $u$. The binary suffix tree can be built in $O(|S|)$
time~\cite{CrochemoreIMS06}.
The generalized suffix tree $\mathcal{T}(S_1, S_2)$ of two strings
$S_1$ and $S_2$ is the suffix tree built over $S' = S_1\$_1S_2\$_2$,
where $\$_1,\$_2\notin\Sigma$, such that the leaves are numbered with
a pair (s-index) and for each leaf $(j, l)$ the concatenation of the
edge labels on the path from the root to the leaf is equal to $S_j[l
  .. |S_j| - 1]\$_j$. The index of a leaf $(j, l)$ is the
starting position of $S_j[l .. |S_j| - 1]\$_j$ in $S_1\$_1S_2\$_2$.
We use the notation $\mathcal{B}(S_1, S_2)$ to denote the binary
generalized suffix tree of $S_1$ and $S_2$.

\section{The longest common substring with $k$ mismatches problem}

Let $S_1$ and $S_2$ be two strings with $n = |S_1|$, $m = |S_2|$.
W.l.o.g. we assume that $n \ge m$. Given an integer $k$, let $\phi(i,
j)$ be the length of the longest substring of $S_1$ and $S_2$ ending
at position $i$ and $j$, respectively, such that the two substrings
have Hamming distance at most $k$. Formally, $\phi(i, j)$ is equal to
the largest integer $l \le \min(i, j) + 1$ such that
$$
|\{ 0\le h\le l - 1\ |\ S_1[i - h]\neq S_2[j - h]\}|\le k\,,
$$
for $0\le i < n, 0\le j< m$. The \emph{longest common substring with
  $k$-mismatches} problem consists in, given two strings $S_1$ and
$S_2$ and an integer $k$, finding the length of the longest substrings
of $S_1$ and $S_2$ with Hamming distance at most $k$, i.e.,
$\max_{i,j} \phi(i, j)$.

\setcounter{instr}{0}
\begin{figure}[t]
\begin{center}
\begin{tabular}{|rl|}
\hline
\multicolumn{2}{|l|}{\textsc{k-lcf}$(S_1, S_2, k)$}\\
\ninstr & $n\leftarrow |S_1|$ \\
\ninstr & $m\leftarrow |S_2|$ \\
\ninstr & $\ell\leftarrow 0, r_1\leftarrow 0, r_2\leftarrow 0$ \\
\ninstr & \textbf{for} $d\leftarrow -m + 1$ \textbf{to} $n-1$ \textbf{do} \\
\ninstr & \qquad $i\leftarrow \max(-d, 0) + d$ \\
\ninstr & \qquad $j\leftarrow \max(-d, 0)$ \\
\ninstr & \qquad $Q\leftarrow \emptyset$ \\
\ninstr & \qquad $s\leftarrow 0, \pos\leftarrow 0$ \\
\ninstr & \qquad \textbf{while} $\pos \le \min(n-i, m-j)-1$ \textbf{do} \\
\ninstr & \qquad \qquad \textbf{if} $S_1[i + \pos]\neq S_2[j + \pos]$ \textbf{then} \\
\ninstr & \qquad \qquad \qquad \textbf{if} $|Q| = k$ \textbf{then} \\
\ninstr & \qquad \qquad \qquad \qquad $s\leftarrow \min Q + 1$ \\
\ninstr & \qquad \qquad \qquad \qquad $\textsc{dequeue}(Q)$ \\
\ninstr & \qquad \qquad \qquad \textsc{enqueue}$(Q, \pos)$ \\
\ninstr & \qquad \qquad $\pos\leftarrow \pos + 1$ \\
\ninstr & \qquad \qquad \textbf{if} $\pos - s > \ell$ \textbf{then} \\
\ninstr & \qquad \qquad \qquad $\ell\leftarrow \pos - s$ \\
\ninstr & \qquad \qquad \qquad $r_1\leftarrow i + s$ \\
\ninstr & \qquad \qquad \qquad $r_2\leftarrow j + s$ \\
\hline
\end{tabular}
\end{center}
\caption{The algorithm to compute the longest common substring up to $k$-mismatches of two strings.}
\label{fig:klcs}
\end{figure}

\section{A practical algorithm for arbitrary $k$}

In this section we present a practical algorithm for the \klcs{k} problem.
By definition, $\phi(i, j)$ is also the length of the longest suffixes
of $S_1[0 .. i]$ and $S_2[0 .. j]$ with Hamming distance at most $k$.
Our algorithm computes all the values $\phi(i, j)$ based on this
alternative formulation.
The idea is to iterate over the $\phi$ matrix diagonal-wise and
compute, for a fixed $(i,j)\in\{(0, 0), (0, 1), \ldots,
(0,m-1)\}\cup\{(1,0), (2,0), \ldots, (n-1,0)\}$, the values $\phi(i +
\pos, j + \pos)$, for $0\le \pos < \min(n-i, m-j)$, i.e., the diagonal starting
at $(i,j)$, in $O(m)$ time. Let $Q$ be an (empty) queue data structure
and $s = 0$, for a given pair $(i,j)$. The algorithm iterates over $\pos$
maintaining the invariant that $\pos - s$ is the length of the longest
common suffix of $S_1[i .. i + \pos - 1]$ and $S_2[j .. j + \pos - 1]$ up to
$k$-mismatches, i.e., $\pos - s = \phi(i + \pos - 1, j + \pos - 1)$, and that
$Q$ contains exactly the positions in $S_1$ of the mismatches between
$S_1[i + s .. i + \pos - 1]$ and $S_2[j + s .. j + \pos - 1]$ with the order
of elements in the queue matching their natural order.

At the beginning the invariant holds since $Q$ is empty, $\pos - s =
0$ and $S_1[i + s .. i + \pos - 1] = S_2[j + s .. j + \pos - 1] = \varepsilon$.
Suppose that the invariant holds up to position $\pos$. If $S_1[i + \pos] =
S_2[j + \pos]$ then the invariant trivially holds also for $\pos+1$ with $s' =
s$ and $Q' = Q$. Otherwise, we have a mismatch between $S_1[i + \pos]$
and $S_2[j + \pos]$. If $|Q| < k$, then the invariant also holds for $\pos+1$
with $s' = s$ and $Q'$ equal to $Q$ after an $\textsc{enqueue}(Q,
\pos)$ operation. Instead, if $|Q| = k$, the pair of suffixes $S_1[i + r .. i
+ \pos]$ and $S_2[j + r .. j + \pos]$, for $r = s, \ldots, \min Q$, match
with $k+1$ mismatches and $r = \min Q + 1$ is the minimum position for
which the corresponding suffixes match with $k$ mismatches. Hence, in
this case the invariant also holds for $\pos+1$ with $s' = \min Q + 1$
and $Q'$ equal to $Q$ after a $\textsc{dequeue}$ operation followed by
an $\textsc{enqueue}(Q, \pos)$ operation.

The algorithm maintains the largest length found up to the current
iteration and the starting positions of the corresponding substrings
in $S_1$ and $S_2$, such that the position in $S_1$ is minimal, in
three integers $\ell$, $r_1$, and $r_2$. Each time $\pos - s > \ell$
it updates their values accordingly. The code of the algorithm is
shown in Figure~\ref{fig:klcs}. The time complexity of one iteration
of the algorithm is $O(1)$ if the queue operations take constant time,
which yields $O(m)$ time for a fixed $i$ and $O(nm)$ time in total.
The space complexity is $O(k)$, as the queue contains at most $k$
elements at any iteration.

For scanning one diagonal of $\phi$, the algorithm needs time that
  is proportional to the length of the diagonal. This can be improved
  such that the time requirement becomes proportional to the number of
  mismatches along the diagonal, by using the well-known technique
  that performs $LCA$ queries on the generalized suffix tree of $S_1$
  and $S_2$ to find, in constant time, how far the next mismatch is
  from the current one~\cite{LandauV86}. This gives an algorithm
  for the \klcs{k} problem that runs in time proportional to the
  number of pairs $(i,j)$ such that $S_1[i] \neq S_2[j]$.

\emph{Constant-space variant:} the algorithm can also be modified to use $O(1)$ space at the price of
a constant factor in the running time. We replace the queue with one
integer $q$, encoding the number of mismatches (number of elements in
the queue). The \textsc{dequeue} and \textsc{enqueue} operations then
become $q\leftarrow q - 1$ and $q\leftarrow q + 1$, respectively. The
update of $s$ requires the computation of $\min Q + 1$, which, by
definition, is equal to the smallest position $s' > s$ such that
$S_1[i + s' - 1]\neq S_2[j + s' - 1]$. To this end, we simply scan
$S_1$ and $S_2$ from position $i + s$ and $j + s$, respectively, until
we find a mismatch. As each symbol of $S_1$ and $S_2$ is looked up at most twice, the
time complexity does not change. In practice, using an explicit queue
is preferable, as it allows one to avoid rescanning the already
scanned parts of the strings.

\emph{Matching statistics with $k$ mismatches:} finally, we describe how to compute the matching statistics with $k$
mismatches of $S_2$ with respect to $S_1$. The matching statistics
with $k$ mismatches of $S_2$ w.r.t. $S_1$ is an array $ms_k$ of $m$
integers such that $ms_k[i]$ is the length of the longest prefix of
$S_2[i .. m - 1]$ that matches a substring of $S_1$ with at most $k$
mismatches, for $i = 0, \ldots, m-1$. Using the algorithm described
above, the array $ms_k$ can be computed in $O(n m)$ time and $O(m)$
space as follows: first, we initialize each slot of $ms_k$ to $0$;
then, we run our algorithm on $S_1^r$ and $S_2^r$, i.e., on the
reverse of the strings $S_1$ and $S_2$, and for each computed cell
$\phi(i, j)$ we set $ms_k[m - 1 - j] = \max(ms_k[m - 1 - j], \phi(i,
j))$. At the end of the procedure we thus have $ms_k[m - 1 - j] =
\max_i \phi(i, j)$, for $0\le j < m$.
The correctness of this procedure follows by observing that i) a
suffix of $S^r[0 .. i]$ is the reverse of a prefix of $S[|S| - 1 - i
  .. |S| - 1]$, for any string $S$ and $0\le i < |S|$, and ii)
$\phi(i, j)$ is the length of the longest suffixes of $S_1^r[0 .. i]$
and $S^r_2[0 .. j]$ with Hamming distance at most $k$. Hence, $\max_i
\phi(i, j)$ is the length of the longest prefix of $S_2[m - 1 - j .. m
  - 1]$ that matches a substring of $S_1$ with at most $k$ mismatches.

Note that the $\phi$ matrix for $S_1$ and $S_2$ immediately gives a
dual matching statistics, where $ms_k[i]$ is defined as the length of
the longest suffix of $S_2[0 .. i]$ that matches a substring of $S_1$
with a most $k$ mismatches. In practical applications this alternative
matching statistics could be equally good.

\section{Longest common substring with $1$ mismatch}

In this section we describe an algorithm that solves the \klcs{1}
problem. We first introduce some necessary technical definitions.
Given a string $S$, a pair of substrings $\SP$ of $S$ is a repeated
pair if $S[p_1 .. q_1] = S[p_2 .. q_2]$. A repeated pair $\SP$ is
left-maximal (right-maximal) if $S[p_1 - 1]\neq S[p_2 - 1]$ ($S[q_1 +
  1]\neq S[q_2 + 1]$).
Given a string $S$, a repeat is a substring of $S$ that corresponds to
a repeated pair. A repeat $w$ of $S$ is left-maximal (right-maximal)
if there exists a left-maximal (right-maximal) repeated pair $\SP$
such that $S[p_1 .. q_1] = S[p_2 .. q_2] = w$. Let $*$ be the don't
care symbol, i.e., a symbol that matches any symbol of $\Sigma$. A
$k$-repeat of $S$ is a string of the form $u *^k v$ that matches more
than one substring of $S$, where $u,v\in\Sigma^*$ and $k > 0$. A longest
$k$-repeat is a $k$-repeat of maximum length.
A necessary condition for a $k$-repeat $u *^k v$ to be
longest is that, for each pair $\SP$ of substrings matching the
repeat, $((p_1, p_1 + |u| - 1), (p_2, p_2 + |u| - 1))$ is a
left-maximal repeated pair and $((p_1 + |u| + k, q_1), (p_2 + |u| + k,
q_2))$ is a right-maximal repeated pair.

The idea is to reduce the \klcs{1} problem to the one of computing
the longest $1$-repeats of $\bar{S} = S_1\$_1S_2\$_2$ that occur in
  both $S_1$ and $S_2$, where $\$_1,\$_2$ are two symbols not in
  $\Sigma$. Let $\ell = \max_{i,j}\phi(i, j)$ for $k = 1$, and let
$i', j'$ be such that $\phi(i', j') = \ell$. Consider the strings $A_1
= S_1[i' - \ell + 1 .. i']$ and $A_2 = S_2[j' - \ell + 1 .. j']$. It
is not hard to see that the string $A_1[0 .. p - 1] * A_1[p + 1 ..
  \ell - 1]$ is a longest $1$-repeat of $\bar{S}$ that occurs in both
$S_1$ and $S_2$, where either $A_1 = A_2$ and $0\le p\le \ell - 1$ or
$A_1 \neq A_2$ and $p$ is the position corresponding to the single
mismatch between $A_1$ and $A_2$.

To this end, we use a modified version of the algorithm
\textsc{all-longest-k-repeats} by Crochemore et al. to find the
longest $k$-repeats of a string~\cite{CrochemoreIMS06}. The idea is to
run this algorithm on the string $\bar{S}$ with $k = 1$. With this
input, the original algorithm reports all the longest $1$-repeats of
$\bar{S}$. To solve our problem we need to add the constraint that the
$1$-repeats must occur in both $S_1$ and $S_2$. As the longest such
repeats can be shorter than the unconstrained longest $1$-repeats of
$\bar{S}$, the \textsc{all-longest-k-repeats} algorithm must be
modified accordingly.

The \textsc{all-longest-k-repeats} algorithm is structured in the following steps:
\begin{enumerate}
\item build the suffix tree $\mathcal{T}(S)$ of $S$ and compute the
  ordering $no$ of the leaves induced by a depth-first visit; build
  the binary suffix tree $\mathcal{B}(S^r)$ of $S^r$ and associate to
  each leaf $u$ with index $i$ a list $\mathcal{A}_u$ equal to $\{
  no(\bar{i}) \}$, if $i \ge k$, and to $\emptyset$ otherwise, where
  $\bar{i} = |S| - i + k$; $\gamma\leftarrow 0$
\item \textbf{for} $u\in\mathcal{B}(S^r)$ in depth-first order with children $u_1$ and $u_2$ \textbf{do}
\begin{enumerate}
\item $\textsc{find-longest}(\mathcal{A}_{u_1}, \mathcal{A}_{u_2}, depth(u) + k, \gamma)$
\item $\mathcal{A}_u\leftarrow \textsc{merge}(\mathcal{A}_{u_1}, \mathcal{A}_{u_2})$
\end{enumerate}
\end{enumerate}
where \textsc{merge}$(L_1, L_2)$ merges two lists $L_1, L_2$, and \textsc{find-longest} is defined as follows:
\begin{center}
\setcounter{instr}{0}
\begin{tabular}{ll}
\multicolumn{2}{l}{$\textsc{find-longest}(L_1, L_2, l, \gamma)$}\\
\ninstr & $(i_1, i_2)\leftarrow (1, 2)$ \\
\ninstr & \textbf{if} $|L_1| > |L_2|$ \textbf{then} $(i_1, i_2)\leftarrow (2, 1)$ \\
\ninstr & \textbf{for} $p\in L_{i_1}$ in ascending order \textbf{do} \\
\ninstr & \qquad $q\leftarrow \max\{j\in L_{i_2}\ |\ j\le p\}, r\leftarrow \min\{j\in L_{i_2}\ |\ j > p\}$ \\
\ninstr & \qquad $v_{pq} = LCA(no^{-1}(p), no^{-1}(q)), v_{pr} = LCA(no^{-1}(p), no^{-1}(r))$ \\
\ninstr & \qquad $\gamma\leftarrow \max(\gamma, l + \max(depth(v_{pq}), depth(v_{pr})))$ \\
\end{tabular}
\end{center}
where the $LCA$ queries are performed on $\mathcal{T}(S)$. At the end
of the algorithm the value of $\gamma$ is the length of the longest
$k$-repeat(s) of $S$. If the lists $L_1, L_2$ are implemented using
AVL-trees, the time complexity of the \textsc{merge} and
\textsc{find-longest} procedures is $O(m\log(n / m))$~\cite{BrownT79},
where $m = \min(|L_1|,|L_2|), n = \max(|L_1|,|L_2|)$, and the
algorithm can be proved to run in $O(|S|\log |S|)$ time. The main
property on which the algorithm is based is the following Lemma:
\begin{lemma}
  Let $u, v, w$ be leaves in the suffix tree of $S$ with corresponding
  depth-first ordering of leaves $no$. If $no(u) < no(v) < no(w)$ or
  $no(w) < no(v) < no(u)$ then $depth(LCA(u, v))\ge depth(LCA(u, w))$.
\end{lemma}
Let $L(u)$ be the list containing the integer $\bar{i}$ for each leaf
with index $i$ in the subtree of node $u$ of $\mathcal{B}(S^r)$. The
idea is to iterate over all the left-maximal repeats of $S$ using
$\mathcal{B}(S^r)$ and for each pair $(p_1, p_2)$ of indexes in $L(u)$
of such a repeat $u$ compute the right-maximal repeat starting at
position $p_1$ and $p_2$ using a LCA query on $\mathcal{T}(S)$. It
turns out, by the above Lemma, that, for a given index $p\in L(u)$, it
is enough to check the pairs $(p, q)$ and $(p, r)$ where $q$ and $r$
are the indexes of the closest leaves to leaf $p$ in $\mathcal{T}(S)$,
with respect to the ordering $no$, such that $q,r\in L(u)$.

Our modification consists in the following: we replace
$\mathcal{T}(S)$ with the generalized suffix tree of $S_1$ and $S_2$
and $\mathcal{B}(S^r)$ with the binary generalized suffix tree of
$S_1^r$ and $S_2^r$. Let $L_j(u)$ be a list containing the integer
$\bar{i}$, for each leaf with s-index $(j, l)$ and index $i$ in the
subtree of node $u$ of $\mathcal{B}(S_1^r,S_2^r)$, provided that $l \ge
k$, for $j=1,2$. The condition $l \ge k$ ensures that the
occurrence of $u$ in $S_j$ ending at position $|S_j| - 1 - l$ can be
extended by $k$ don't care symbols to the right, as otherwise there
can be no $k$-repeat with left part equal to the reverse of $u$ label
matching a prefix of $S_j[|S_j| - 1 - l - depth(u) + 1 .. |S_j| - 1]$.
Our goal is to iterate over pairs in $L_1(u)\times L_2(u)$ only by
computing, for a given index $p\in L_1(u)$, the indexes $q$ and $r$ of
the closest leaves to leaf $p$ in $\mathcal{T}(S_1, S_2)$, with
respect to the ordering $no$, such that $q,r\in L_2(u)$, and
\emph{vice versa} if $p\in L_2(u)$. To accomplish this, it is enough
to associate to each leaf $u$ of $\mathcal{B}(S_1^r,S_2^r)$ with s-index
$(j,l)$ and index $i$ two lists, $\mathcal{A}^1_u$ and
$\mathcal{A}^2_u$: if $l < k$ the lists are empty; otherwise, if
$j = 1$ then $\mathcal{A}^1_u = \{ no(\bar{i}) \}$ and
$\mathcal{A}^2_u = \emptyset$, and \emph{vice versa} if $j = 2$. Then,
we change the operations in the second step of the algorithm
as follows:
\begin{enumerate}[(a)]
\item $\textsc{find-longest}(\mathcal{A}^1_{u_1}, \mathcal{A}^2_{u_2}, depth(u) + k, \gamma)$
\item $\textsc{find-longest}(\mathcal{A}^2_{u_1}, \mathcal{A}^1_{u_2}, depth(u) + k, \gamma)$
\item $\mathcal{A}^1_u\leftarrow \textsc{merge}(\mathcal{A}^1_{u_1}, \mathcal{A}^1_{u_2})$
\item $\mathcal{A}^2_u\leftarrow \textsc{merge}(\mathcal{A}^2_{u_1}, \mathcal{A}^2_{u_2})$
\end{enumerate}
In this way we iterate only over pairs $\SP$ of $\bar{S}$ matching a
$1$-repeat and such that $0\le p_1, q_1\le |S_1| - 1$ and $|S_1| +
1\le p_2,q_2\le |S_1| + |S_2|$, or \emph{vice versa}. At the end of
the algorithm the value of $\gamma$ is the length of the longest
$k$-repeat(s) of $\bar{S}$ that occur in both $S_1$ and $S_2$.

We now prove that the time complexity of steps a, b, c, and d is
$O(m\log(n / m))$, where $m =
\min(|\mathcal{A}_{u_1}|,|\mathcal{A}_{u_2}|), n =
\max(|\mathcal{A}_{u_1}|,|\mathcal{A}_{u_2}|)$, i.e., there is only a
constant overhead compared to the original algorithm. Suppose w.l.o.g.
that $m=|\mathcal{A}_{u_1}|, n=|\mathcal{A}_{u_2}|$ and let $m_i =
|\mathcal{A}^i_{u_1}|$ and $n_j = |\mathcal{A}^j_{u_2}|$, for $1\le
i,j\le 2$. Note that $m\ge m_1 + m_2$ and $n\ge n_1 + n_2$. Step a, b,
c, or d takes i) $O(m_i\log(n_j / m_i)) = O(m_i\log(n / m_i))$ time,
if $m_i\le n_j$; ii) $O(n_j\log(m_i / n_j)) = O(n_j\log(n / n_j))$
time otherwise, where $n_j \le m_i$. We show that $a\log(n / a) \le
m\log(n / m)$ for any $1\le a\le m$. This inequality can be written as
$ \frac{f(m) - f(a)}{m - a}\le \log n$ where $f(x) = x\log x$. We have
$f'(x) = \log x$ and, by the mean value theorem, there exists $c\in
(a, m)$ such that $\frac{f(m) - f(a)}{m - a} = \log c\le \log m\le
\log n$.

The total time complexity of our algorithm for the \klcs{1} problem is
thus $O((n + m)\log(n + m))$. Assuming $m\le n$, we can reduce it to
$O(n\log m)$ by partitioning $S_1$ into overlapping substrings of
length $2m$ such that the overlap between two consecutive substrings
is of length $m$, and running the algorithm on each substring and
$S_2$. Formally, we run the algorithm on $S_1[m\cdot i .. \min(m\cdot
  i + 2m, n) - 1]$ and $S_2$ and obtain a value $\gamma_i$, for $0\le
i< \lceil n / m\rceil$. Then, $\ell = \max_i \gamma_i$. The time
complexity of this algorithm is $O((n/m) m\log m) = O(n\log m)$.

\section{Acknowledgments}

We thank the anonymous reviewers for helpful comments.

\bibliographystyle{abbrv}
\bibliography{klcf}

\begin{thebibliography}{10}

\bibitem{Babenko11}
M.~A. Babenko and T.~A. Starikovskaya.
\newblock Computing the longest common substring with one mismatch.
\newblock {\em Problems of Information Transmission}, 47(1):28--33, 2011.

\bibitem{BenderF00}
M.~A. Bender and M.~Farach-Colton.
\newblock The {LCA} problem revisited.
\newblock In {\em LATIN}, pages 88--94, 2000.

\bibitem{BrownT79}
M.~R. Brown and R.~E. Tarjan.
\newblock A fast merging algorithm.
\newblock {\em J. {ACM}}, 26(2):211--226, 1979.

\bibitem{ChangL94}
W.~I. Chang and E.~L. Lawler.
\newblock Sublinear approximate string matching and biological applications.
\newblock {\em Algorithmica}, 12(4/5):327--344, 1994.

\bibitem{CrochemoreIMS06}
M.~Crochemore, C.~S. Iliopoulos, M.~Mohamed, and M.-F. Sagot.
\newblock Longest repeats with a block of {\it k} don't cares.
\newblock {\em Theor. Comput. Sci.}, 362(1-3):248--254, 2006.

\bibitem{Gusfield1997}
D.~Gusfield.
\newblock {\em Algorithms on Strings, Trees, and Sequences - Computer Science
  and Computational Biology}.
\newblock Cambridge University Press, 1997.

\bibitem{KociumakaSV14}
T.~Kociumaka, T.~A. Starikovskaya, and H.~W. Vildh{\o}j.
\newblock Sublinear space algorithms for the longest common substring problem.
\newblock In {\em ESA}, pages 605--617, 2014.

\bibitem{LandauV86}
G.~M. Landau and U.~Vishkin.
\newblock Introducing efficient parallelism into approximate string matching
  and a new serial algorithm.
\newblock In {\em STOC}, pages 220--230, 1986.

\bibitem{LeimeisterM14}
C.-A. Leimeister and B.~Morgenstern.
\newblock kmacs: the {\it k}-mismatch average common substring approach to
  alignment-free sequence comparison.
\newblock {\em Bioinformatics}, 30(14):2000--2008, 2014.

\bibitem{SmithW81}
T.~F. Smith and M.~S. Waterman.
\newblock Identification of common molecular subsequences.
\newblock {\em Journal of Molecular Biology}, 147(1):195--197, 1981.

\bibitem{StarikovskayaV13}
T.~A. Starikovskaya and H.~W. Vildh{\o}j.
\newblock Time-space trade-offs for the longest common substring problem.
\newblock In {\em CPM}, pages 223--234, 2013.

\end{thebibliography}

\end{document}